\documentclass[prb,twocolumn,superscriptaddress,showpacs,citeautoscript,floatfix]{revtex4}
\usepackage{amsmath}
\usepackage{amssymb}
\usepackage{graphicx}

\begin{document}

\title{
Thermoelectric properties of a quantum dot coupled to magnetic leads by Rashba spin-orbit interaction}

\author{\L{}ukasz Karwacki}
\email{karwacki@ifmpan.poznan.pl}
\author{J\'{o}zef Barna\'{s}}
\affiliation{Faculty of Physics, Adam Mickiewicz University, ul. Umultowska 85, 61-614 Pozna\'{n},
Poland}
\affiliation{Institute of Molecular Physics, Polish Academy of Sciences, ul. M. Smoluchowskiego 17, 60-179 Pozna\'{n}, Poland}

\begin{abstract}
We consider a single-level quantum dot coupled to two leads which are ferromagnetic in general. Apart from tunneling processes conserving electron spin, we also include processes associated with spin-flip of tunneling electrons, which appear due to Rashba spin-orbit coupling. Charge and heat currents are calculated within the  non-equilibrium Green's function technique. When the electrodes are half-metallic (fully spin polarized), the Rashba spin-orbit coupling leads to Fano-like interference effects, which result in an enhanced thermoelectric response. It is also  shown that such a system can operate as efficient heat engine. Furthermore, the interplay  of Rashba spin-orbit coupling and Zeeman splitting due to an external magnetic field is shown to allow control over such parameters of the heat engine as the power and efficiency.
\end{abstract}

\pacs{07.20.Pe,84.60.Rb,75.70.Tj,81.07.Ta}
\maketitle

\section{Introduction}
\label{sec:intro}

Thermoelectric properties of nanoscale systems have been the subject of many recent studies in condensed matter physics~\cite{Hicks,Dresselhaus}. It has been shown theoretically that high thermoelectric figure of merit, which is a measure of thermoelectric efficiency, can be obtained in zero-dimensional (0D) systems with discrete density of states (DOS)~\cite{Mahan}, such as quantum dots (QDs) or molecules. The discrete  DOS of quantum dots, combined with electric tunability of their energy levels, leads to strong energy filtering of particles. Furthermore, quantum dots can operate in different transport regimes, from a weakly coupled system to a strongly correlated one. Each regime displays distinct behavior with characteristic energy scales. One of the common characteristic features for both regimes is the sign alternation of Seebeck coefficient with a gate voltage applied to the dot, which has been verified experimentally~\cite{Staring,Scheibner,Svensson,Svilans}. This effect results from strongly bipolar transport in quantum dots, where tuning an energy level of the dot (with gate voltage) around the Fermi level of the electrodes filters either holes or electrons. Similar effect can be observed when the time-reversal symmetry is broken due to either external magnetic field or ferromagnetic electrodes. The quantum dots can then filter spin-up or spin-down electrons, which results in spin-dependent transport and spin-dependent thermoelectric effects~\cite{Krawiec,Swirkowicz,Rejec,Weymann}.

Another important property of quantum dots, extensively studied theoretically, is a large impact of quantum interference effects on electronic transport in different regimes. One of such phenomena, known as Fano effect~\cite{Fano}, can occur when one of the discrete transport channels is coupled to continuum of states in the electrodes, while the second one is decoupled from the electrodes but remains active due to interaction with the other channel. It has been predicted that such an effect in multiple quantum dot structures can lead to characteristic antiresonances in electrical conductance, and simultaneously  can also lead to enhanced spin and charge Seebeck coefficients~\cite{Wierzbicki,Trocha2012,GarciaSuarez,KarwackiPRB,Wojcik,RamosAndrade}.

The most common quantum dots are based on two-dimensional electron gas (2DEG) confined at the interface between two semiconductors, which makes  it relatively easy to control properties of the dots with gate voltages. Alternatively, the dots can be created in one-dimensional (1D) structures such as semiconductor nanowires or carbon nanotubes (CNT)~\cite{Bjork,JarilloHerrero,Sapmaz}.
One of the effects that can inevitably arise in such systems due to inversion-symmetry breaking at interfaces or due to curvature of CNT is Rashba spin-orbit coupling (RSOC), which also can be controlled by electric means~\cite{Bychkov,Bercioux,Kuemmeth,Steele}. The spin-orbit coupling can limit spin coherence due to spin-mixing of transport channels~\cite{Mireles,Orellana,Stefanski}. However, the spin reversal due to spin-orbit coupling can be also used to induce quantum interference phenomena. For instance, interference effects in mesoscopic structures with Rashba spin-orbit coupling have been proposed in ring interferometers, where Aharonov-Bohm effect and spin-dependent phase shift between different paths traversed by spin-$\uparrow$ and spin-$\downarrow$ electrons appear~\cite{Nitta,Sun}. Theoretical studies of such structures indicate some enhancement of the thermoelectric figure of merit and possibility of pure spin current generation~\cite{Liu,LiuJAP}.

Although a lot of the above properties have been already studied, mostly in the linear response regime, there have been recently many proposals of quantum-dot-based heat engines, where one needs to go beyond the linear response limit. The heat engine  based on a single-level quantum dot coupled to two metallic reservoirs
has been theoretically predicted to reach either the Curzon-Ahlborn efficiency, i.e. the efficiency of the engine at maximum produced power, or even the Carnot efficiency when a strong coupling between the heat and particle fluxes exists~\cite{Curzon, EspositoPRL2009,EspositoPRL2010,EspositoPRE2010,Nakpathomkun2010,EspositoPRE2012,Szukiewicz}. Although this condition occurs for very weakly coupled quantum dots, the possibility of achieving high thermodynamic efficiency in quantum dot systems has been recently verified experimentally~\cite{Josefsson}. More complex  heat engines, based on multiple quantum dots and multiple electrodes,  have been proposed as well~\cite{Wohlman,Thierschmann, Sothmann,Bergenfeldt,Mazza}. Conversely, also mesoscopic refrigeration schemes have been proposed~\cite{Timofeev,Venturelli,PekolaPRB2014}, paving way to studies on quantum heat transport and its relation with information in the form of Maxwell's demon~\cite{PekolaNPhys2015,Strasberg}.

Here we show that the quantum dot with Rashba spin-orbit coupling can operate as a heat engine with the  efficiency that can be controlled not only by  position of the dot's energy level and external magnetic field, but also by strength of the Rashba spin-orbit coupling. Moreover, we show that when the quantum dot is coupled to half-metallic electrodes, the Rashba spin-orbit coupling gives rise to the Fano-like interference effect. This effect leads to higher thermoelectric parameters and enhanced efficiency of the heat engine. More complex effects appear when external magnetic field is applied to the system.

The paper is organized as follows. Section~\ref{sec:theo} contains description of the model and of the quantities being considered (charge and heat currents). Basic information on the power produced by a heat engine and the  corresponding efficiency is also presented there. Section~\ref{sec:numres} presents numerical results obtained for the quantities introduced in Sec. II. Short summary of the paper is presented in Sec.~\ref{sec:concl}.

\section{Theoretical description}
\label{sec:theo}

\subsection{Model}
\label{sec:model}

\begin{figure}
\includegraphics[width=0.9\columnwidth]{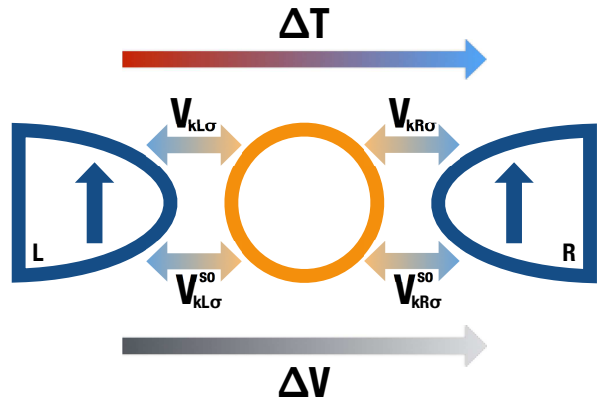}
\caption{Schematic presentation of the quantum dot coupled to two ferromagnetic electrodes. The coupling parameter $V_{\mathbf{k}\beta\sigma}$ ($\beta=L,R$) represents spin-conserving tunneling process, while the parameter $V_{\mathbf{k}\beta\sigma}^{\text{so}}$ represents spin-nonconserving tunneling process due to Rashba spin-orbit coupling. Temperature and electrostatic potential of the left electrode are shifted by $\Delta T$ and $\Delta V$ in comparison to the right electrode.
}
\label{fig:fig1}
\end{figure}

The system under consideration is presented schematically in Fig.~\ref{fig:fig1}.
The quantum dot is coupled to two electrodes by direct (spin conserving) tunneling and tunneling with Rashba interaction (spin non-conserving). The system can be described by the following Hamiltonian:
\begin{equation}
\label{eq:ham0}
\hat{\text{H}}=\hat{\text{H}}_{\text{e}}+\hat{\text{H}}_{\text{qd}}+\hat{\text{H}}_{\text{t}}^0+\hat{\text{H}}_{\text{t}}^{\text{so}}\,,
\end{equation}
where the first term,
\begin{equation}
\hat{\text{H}}_{\text{e}}=\sum_{\mathbf{k}\beta\sigma}\varepsilon_{\mathbf{k}\beta\sigma}c_{\mathbf{k}\beta\sigma}^\dagger c_{\mathbf{k}\beta\sigma}\,,
\end{equation}
describes the left ($\beta=L$) and right ($\beta=R$) electrodes, which are ferromagnetic in a general case. Magnetic moments of the electrodes are assumed to be collinear, and orientation of the moments  determines a quantization axis for the system.

The second term in Hamiltonian~(\ref{eq:ham0}) stands for the single-level quantum dot,
\begin{equation}
\hat{\text{H}}_{\text{qd}} = \sum_{\sigma}\varepsilon_{\sigma}d_{\sigma}^\dagger d_{\sigma} + U\hat{n}_{\uparrow}\hat{n}_{\downarrow}\,,
\end{equation}
where $\varepsilon_\sigma=\varepsilon_d+\hat{\sigma}g\mu_BB/2$, with $\varepsilon_d$ being the bare dot's level, $\hat{\sigma}$ defined as $\hat{\sigma}=1 (-1)$ for $\sigma=\uparrow(\downarrow)$, $B$ denoting an external magnetic field, and $g$ and $\mu_B$ standing for the Land\'{e} factor and Bohr magneton, respectively. In turn, $U$ in Eq.~(3) is the Coulomb correlation parameter, while $\hat{n}_\uparrow (\hat{n}_\downarrow )$ is the occupation operator for spin-up (spin-down) electrons.

The last two terms in Eq.~(1) describe electron tunneling between the electrodes and quantum dot.  One of them, $\hat{\textrm{H}}_{\textrm{t}}^0$,  conserves electron spin in the tunneling processes,
\begin{equation}
\hat{\text{H}}_{\text{t}}^0=\sum_{\textbf{k}\beta\sigma}V_{\textbf{k}\beta\sigma}c_{\textbf{k}\beta\sigma}^{\dagger}d_{\sigma}+{\rm H.c.}\,,
\end{equation}
while the second one, $\hat{\text{H}}_{\text{t}}^{\text{so}}$, is of the form
\begin{equation}
\label{eq:HamSO}
\hat{\text{H}}_{\text{t}}^{\text{so}}=-\sum_{\textbf{k}\beta\sigma}\left[V_{\textbf{k}\beta\overline{\sigma}}^{\text{so}}c_{\textbf{k}\beta\overline{\sigma}}^{\dagger}
(i\hat{\sigma}_{x})_{\sigma\overline{\sigma}}d_{\sigma}\right]+{\rm H.c.}\,,
\end{equation}
and flips the electron spin in the tunneling processes due to Rashba spin-orbit coupling. In the expression above  $\hat{\sigma}_x$ denotes the appropriate Pauli matrix. When the leads are half-metallic, the above-introduced model is equivalent to a spinless two-level quantum dot model.~\cite{Kascheyevs,KarwackiJPCM}

\subsection{Currents and heat engine}
\label{sec:currents}

The electric current $j_e$ flowing in the biased system from left to right can be described by the formula
\begin{equation}
\label{eq:elcurr}
j_{e}=\frac{e}{\hbar}\int\frac{\textrm{d}\varepsilon}{2\pi}\left[f_L(\varepsilon)-f_R(\varepsilon) \right]T(\varepsilon)\,,
\end{equation}
where $e$ denotes the electron charge ($e<0$),  $f_{L(R)}$ is the Fermi-Dirac distribution in the left (right) electrode, while
$T(\varepsilon)$ is the total transmission function, $T(\varepsilon)=T_\uparrow(\varepsilon)+T_\downarrow(\varepsilon)$,
whose explicit form will be derived in the next subsection. The spin dependent transmission $T_{\uparrow (\downarrow )}(\varepsilon)$ is defined as the total transmission from the spin-$\sigma$ channel of one electrode to both spin channels in the second electrode.

Since the dot is coupled, in general, to ferromagnetic electrodes,
the charge current may be accompanied with a spin current. However, we assume no spin accumulation in the leads (no spin voltage and no spin thermoelectric effects), and therefore we do not consider the spin currents.

The charge current is also associated with energy flow, and the corresponding energy current is given by the formula
\begin{equation}
\label{eq:encurr}
j_{E}=\frac{1}{\hbar}\int\frac{\textrm{d}\varepsilon}{2\pi}\varepsilon\left[f_L(\varepsilon)-f_R(\varepsilon) \right]T(\varepsilon)\,.
\end{equation}
The energy current is conserved, but the associated heat current  is not conserved in nonequilibrium situations. According to the second law of thermodynamics, the energy increase of the dot is equal to the heat flowed into it and the work done on it.
The heat current flowing from the left electrode can be written as
\begin{equation}
\label{eq:hcurr}
j_{h}^L=\frac{1}{\hbar}\int\frac{\textrm{d}\varepsilon}{2\pi}(\varepsilon-\mu_L)\left[f_L(\varepsilon)-f_R(\varepsilon) \right]T (\varepsilon)\,,
\end{equation}
where  $\mu_{L}$ is the electrochemical potential of the left electrode. Similar formula holds for the heat flowing from the right electrode to the dot. These heat currents are generally different.
The heat current is conserved only in quasi-equilibrium state (infinitesimally small deviation from equilibrium).

In the linear response regime (quasi-equilibrium situation), the charge  and heat currents driven by small bias voltage $\delta V$ and temperature difference $\delta T$ can be written in the following form:
\begin{align}
\renewcommand*{\arraystretch}{1.5}
\begin{bmatrix}j_e \\ j_{h}\end{bmatrix} =
\begin{bmatrix} e^2 L_{0} &  \dfrac{e}{T}L_{1}\\
eL_1 &   \dfrac{1}{T}L_{2} \end{bmatrix}
\begin{bmatrix}\delta V  \\  \delta T\end{bmatrix}\,,
\end{align}
where
\begin{equation}
L_{i}=\frac{1}{\hbar}\int\frac{\textrm{d}\varepsilon}{2\pi}(\varepsilon-\mu)^i\left(-\frac{\partial f_0}{\partial \varepsilon}\right)T(\varepsilon)\,
\end{equation}
for $i=0,1,2$. Here, $f_0$ is the Fermi-Dirac distribution function in equilibrium (corresponding to the chemical potential $\mu$). Note, according to our definitions, $\delta V =V_L-V_R = (\mu_L-\mu_R)/e$ and $\delta T =T_L-T_R$.
The electrical conductance, $G$, can be then calculated as $G=e^2L_{0}$, while the thermopower, $S$, is given by the formula $S=-\frac{\delta V}{\delta T}=\frac{1}{eT}\frac{L_{1}}{L_{0}}$. The linear-response $G$ and $S$ determine the  corresponding power factor $P_0$ as $P_0=GS^2$.

When the system is supposed to work as a heat engine, the linear response regime is then not sufficient and one needs to go beyond this limit. In other words,  $\delta V$ and $\delta T$ should be replaced by a finite (not small) $\Delta V$ and $\Delta T$, where transport characteristics are nonlinear. The charge and heat currents cannot be then calculated from Eqs.~(9), but instead one should use Eqs.~(6) and (8). Note, the conductance $G=j_e/\Delta V$, the thermopower $S$,
\begin{equation}
S=-\frac{\Delta V}{\Delta T},
\end{equation}
and the corresponding power factor $P_0=GS^2$ depend then on the voltage $\Delta V$.

The work done on the system per unit time is $j_{e}\Delta V$. When the system operates as a heat engine, it generates a finite power,
\begin{equation}
P=-j_e\Delta V\,,
\end{equation}
where $\Delta V$ is the voltage applied to counteract the thermally-induced current.
The maximal power generated by the engine can be described with the following formula~\cite{Benenti}:
\begin{equation}
P_{\textrm{max}}=GV_{\textrm{max}}^2=\frac{1}{4}GV_{b}^2=\frac{1}{4}P_0(\Delta T)^2\,,
\end{equation}
where $V_{\textrm{max}}=V_b/2$ is the voltage for which power is maximal, while $V_{b}=-S\Delta T$ is the stopping (or blocking) voltage.

Efficiency of the heat engine is defined as
\begin{equation}
\eta=\frac{P}{j_h^L}\,.
\end{equation}
The second law of thermodynamics introduces the upper limit on the efficiency in the form of Carnot efficiency,
\begin{equation}
\eta_{\textrm{C}}=\frac{\Delta T}{T}\,,
\end{equation}
where $T$ is the temperature of the hotter (here left) reservoir, $T=T_L$.
Additionally, for realistic heat engines, when one considers device's output at maximal power, a Curzon-Ahlborn efficiency can be introduced~\cite{Curzon},
\begin{equation}
\eta_{\textrm{CA}}=1-\sqrt{1-\eta_{\textrm{C}}}\,,
\end{equation}
which for strongly coupled particle and energy currents in the linear response acquires a finite value, $\eta_{\textrm{CA}}\approx\eta_{\textrm{C}}/2+\mathcal{O}(\eta_{\textrm{C}}^2)$~\cite{EspositoPRL2009}.

\subsection{Method}
\label{sec:method}

To find the charge  and heat currents introduced  above, one needs to know the transmission coefficient $T(\varepsilon)=T_\uparrow(\varepsilon)+T_\downarrow(\varepsilon)$. In the mean field approximation for the Coulomb interaction in the dot, one finds
\begin{equation}
\label{eq:ch2_trans}
T_{\sigma}(\varepsilon)=\operatorname{Tr}\left\lbrace \boldsymbol{\Gamma}_{L\sigma}\mathbf{G}^r(\varepsilon)\boldsymbol{\Gamma}_R\mathbf{G}^a(\varepsilon) \right\rbrace\,
\end{equation}
for $\sigma=\uparrow,\downarrow$.
This transmission coefficient can be understood as coupling of spin $\sigma$ state from the left electrode to both spin states of the right electrode, i.e. it can be decomposed into spin-conserving transmission, and transmission with spin reversal due  to the spin-orbit coupling, i.e. spin-mixing transmission coefficient. The $T_\sigma(\varepsilon)$ coefficient can be equivalently  defined as coupling of both spin states from left electrode to a selected spin state in the right electrode.

The Green's functions can be derived from the Dyson equation,
\begin{equation}
\label{eq:ch2_Dyson}
\mathbf{G}^{\text{r(a)}}=\left[(\mathbf{g}_0^{\text{r(a)}})^{-1}-\boldsymbol{\Sigma}^{\text{r(a)}} \right]^{-1}\,,
\end{equation}
with $\mathbf{g}_0^{\text{r(a)}}$ being the Green's function of the corresponding isolated dot, whose diagonal elements are defined as follows:
\begin{equation}
g_{0\sigma\sigma}^{\text{r(a)}}=\frac{1-n_{\overline{\sigma}}}{\varepsilon-\varepsilon_{\sigma}\pm i0^{+}}+\frac{n_{\overline{\sigma}}}{\varepsilon-\varepsilon_{\sigma}-U\pm i0^{+}}\,,
\end{equation}
where $n_\sigma =\langle \hat{n}_\sigma\rangle$.
The self-energy takes the following form:
\begin{equation}
\boldsymbol{\Sigma}^{\text{r(a)}}=\mp\frac{i}{2}\sum_\beta\boldsymbol{\Gamma}_\beta\,,
\end{equation}
where $\boldsymbol{\Gamma}_\beta=\boldsymbol{\Gamma}_{\beta\uparrow}+\boldsymbol{\Gamma}_{\beta\downarrow}$. The coupling matrices are  defined as
\begin{equation}\label{eq:GammaUp}
\boldsymbol{\Gamma}_{\beta\uparrow}=
\begin{bmatrix}
\Gamma_{\beta\uparrow\uparrow} & -i\sqrt{\Gamma_{\beta\uparrow\uparrow}\Gamma_{\beta\uparrow\uparrow}^{\text{so}}} \\
i\sqrt{\Gamma_{\beta\uparrow\uparrow}\Gamma_{\beta\uparrow\uparrow}^{\text{so}}} & \Gamma_{\beta\uparrow\uparrow}^{\text{so}}
\end{bmatrix}\,
\end{equation}
for spin $\sigma=\uparrow$, and
\begin{equation}\label{eq:GammaDown}
\boldsymbol{\Gamma}_{\beta\downarrow}=
\begin{bmatrix}
\Gamma_{\beta\downarrow\downarrow}^{\text{so}} & i\sqrt{\Gamma_{\beta\downarrow\downarrow}\Gamma_{\beta\downarrow\downarrow}^{\text{so}}} \\
-i\sqrt{\Gamma_{\beta\downarrow\downarrow}\Gamma_{\beta\downarrow\downarrow}^{\text{so}}} & \Gamma_{\beta\downarrow\downarrow}
\end{bmatrix}\,
\end{equation}
for spin $\sigma=\downarrow$.
In the above matrices $\Gamma_{\beta\sigma\sigma}=2\pi\langle |V_{\beta {\mathbf k}\sigma}|^2\rangle \rho_{\beta\sigma}$ and $\Gamma_{\beta\sigma\sigma}^{\text{so}}=2\pi\langle |V_{\beta {\mathbf k}\sigma}^{\text{so}}|^2\rangle \rho_{\beta\sigma}$, with $\rho_{\beta\sigma}$ denoting the spin-dependent density of states in the $\beta$-th lead.  In the following, we introduce the parameters $\Gamma_\beta$ through the relation  $\Gamma_{\beta\sigma\sigma}=(1+\hat{\sigma}p_\beta)\Gamma_\beta$ and the parameter $q$ defined as  $\Gamma_{\beta\sigma\sigma}^{\text{so}}=q\Gamma_{\beta\sigma\sigma}$. The parameter $q$ describes  the relative strength of Rashba spin-orbit coupling. Accordingly, the parameters $\Gamma_L$, $\Gamma_R$, and $q$ will be used to describe coupling of the dot to both leads.

To obtain the dot's mean occupation for spin-$\sigma$ electrons, $n_{\sigma}=-i\int\textrm{d}\varepsilon/2\pi G_{\sigma }^<$, we use the Keldysh formula,
\begin{equation}
\label{eq:ch2_Keldysh}
\mathbf{G}^<=i\mathbf{G}^{\text{r}}\left(f_L\boldsymbol{\Gamma}_L+f_R\boldsymbol{\Gamma}_R \right)\mathbf{G}^{\text{a}}\,,
\end{equation}
where $\mathbf{G}^<$ is the correlation (lesser) Green function.

\section{Numerical results}
\label{sec:numres}

In this section we present some numerical results. The section is divided into three parts: in the first one we consider mean occupation of the dot by spin-$\uparrow$ and spin-$\downarrow$ electrons in the absence and presence of an external magnetic field.  Thermoelectric effects and heat engine in the absence of magnetic field are analyzed in the second part, while the influence of a finite magnetic field is considered in the third part.
In all calculations we assumed symmetrical coupling of the dot to both leads, $\Gamma_L=\Gamma_R=\Gamma$. Apart from this, we assumed  $U=10\Gamma$ and $k_BT=0.5\Gamma$  (unless otherwise specified), where $\Gamma=0.01D$ is used as the energy unit, with $D$ being the leads' half-bandwidth.
Note that the temperatures considered here are much higher than the corresponding Kondo temperature. The problem of Kondo correlations in the model under consideration was investigated elsewhere~\cite{KarwackiJPCM}.

\subsection{Mean occupations and average spin}

\begin{figure}
\includegraphics[width=\columnwidth]{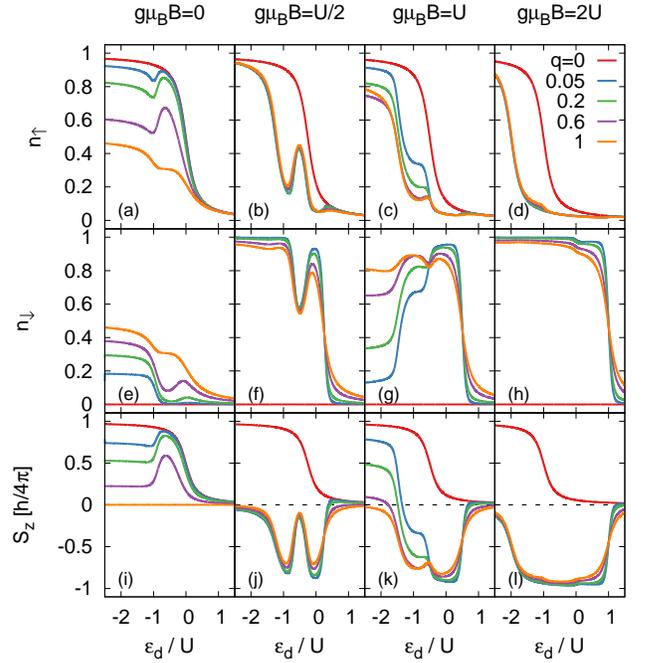}
\caption{Mean occupation of the dot by a spin-$\uparrow$ electron, $n_\uparrow$, [(a)-(d)], by spin-$\downarrow$ electron, $n_\downarrow$, [(e)-(h)], and the average spin of the dot, $S_z/(\hbar/2)\equiv n_\uparrow -n_\downarrow$, [(i)-(l)], presented as a function of the dot's bare energy level, $\varepsilon_d$, and calculated in the linear response regime for indicated values of the Rashba spin-orbit parameter $q$ and Zeeman splitting $g\mu_BB$. Other parameters: $p_L=p_R=1$, $k_BT=0.5\Gamma$, and $U=10\Gamma$. The dashed line indicates $S_z=0$.}
\label{fig:fig2}
\end{figure}

In order to better understand complex behavior of the thermoelectric effects in quantum dots coupled to  half-metallic leads {\it via} tunneling with  Rashba spin-orbit interaction, it is useful to analyze first occupation of the dot by spin-$\uparrow$ and spin-$\downarrow$ electrons as well as the average spin on the dot. All these parameters are shown in Fig.~\ref{fig:fig2} as a function of the quantum dot's energy level and for selected values of the Rashba spin-orbit coupling parameter $q$ and fixed values of magnetic field $g\mu_BB$.

The occupation numbers for spin-$\uparrow$ and spin-$\downarrow$ electrons as well as the dot's average spin $S_z$ in the absence of external magnetic field, $g\mu_BB=0$, and for half-metallic leads, $p_L=p_R=1$, are shown in Figs.~\ref{fig:fig2}(a),~\ref{fig:fig2}(e), and~\ref{fig:fig2}(i), respectively. When the Rashba spin-orbit coupling is absent, i.e. $q=0$, the dot can be occupied only by a single spin-$\uparrow$ electron, which can enter the dot when its energy  is in resonance with the dot's energy level $\varepsilon_d$.
In turn, when the Rashba spin-orbit coupling is nonzero, the quantum dot can also be occupied by a spin-$\downarrow$ electron. Thus, the dot can be then either empty, or singly occupied by a spin-$\uparrow$ or spin-$\downarrow$ electron,  or occupied by two
electrons (spin-$\uparrow$ and spin-$\downarrow$). Note,  the
second electron can enter the dot when its energy overcomes the Coulomb blockade. Due to the appearance of a spin-$\downarrow$ electron in the dot, the average spin $S_z$ of the dot is adequately reduced. Further increase in Rashba spin-orbit coupling leads to a further increase of the dot's mean occupation number for  a spin-$\downarrow$ electron, and thus a to a further decrease in the average spin. In the case of $q=1$, the dot can be occupied equally by spin-$\uparrow$ and spin-$\downarrow$ electrons, which results in  zero average spin, irrespective of the dot's energy level.

The occupation numbers in the presence of a  moderate external magnetic field ($g\mu_BB=U/2$) are shown in  Figs.~\ref{fig:fig2}(b),~\ref{fig:fig2}(f), and~\ref{fig:fig2}(j). For $q=0$ the situation is similar to that in the absence of magnetic field, i.e. only a single spin-$\uparrow$ electron can enter the dot and the average spin of the dot is maximal. Increase in the Rashba spin-orbit coupling (parameter $q$) leads to remarkable changes in the dependence of the occupation numbers $n_\uparrow$ and $n_\downarrow$ on the dot's energy level $\varepsilon_d$. For the assumed  magnetic field, the energy level $\varepsilon_\downarrow$ is at the Fermi level of the leads when $\varepsilon_d=U/4$, while the $\varepsilon_\uparrow$ level is then above the Fermi level. The dot becomes then occupied mainly by a spin-$\downarrow$ electron, and thus the average spin is negative.
Note, that due to a finite temperature a spin-$\uparrow$ electron can enter the dot as well, leading to a small but nonzero value of  $n_\uparrow$.
For $\varepsilon_d=-U/4$, the $\varepsilon_\uparrow$ level is activated in transport and a spin-$\uparrow$ electron can enter the dot as well, decreasing the average spin in the dot and reducing the occupation number for spin-$\downarrow$ electrons. 
Further decrease in energy of the dot's level leads firstly to an increase of the negative average spin and then to a further reduction of the average spin, as the dot's mean occupation numbers $n_\uparrow$ and $n_\downarrow$ become approximately equal when the the Coulomb blockade is overcome.

The mean occupations and average spin in the specific case of  $g\mu_BB=U$, shown in Figs.~\ref{fig:fig2}(c),~\ref{fig:fig2}(g), and~\ref{fig:fig2}(k), respectively, indicate that it is possible to switch the dot's spin by sweeping the dot's level (due to gate voltage) and tuning  strength of the spin-orbit coupling. For $0<q<1$, the average spin on the dot becomes negative when shifting the dot's level $\varepsilon_d$ below $\varepsilon_d=U/2$.
However, for $\varepsilon_d\approx -U/2$, the $\varepsilon_\uparrow$ level crosses the Fermi level and
the occupation of the dot by a spin-$\uparrow$ electron increases. When $\varepsilon_d\approx -3U/2$, the blockade is overcome and double occupancy is possible, giving rise to a further increase in $n_\uparrow$. The average spin becomes then positive. However, for $q=1$, the probability of the occupation by a spin-$\uparrow$ electron is equal to that for spin-$\downarrow$ electron, and the average spin is then zero.

Finally, when $g\mu_BB=2U$, see Figs.~\ref{fig:fig2}(d),~\ref{fig:fig2}(h), and~\ref{fig:fig2}(l), the dot for $q>0$ is mostly occupied by a spin-$\downarrow$ electron in the energy range shown there. For $\varepsilon_d = -2U$ the Coulomb blockade is overcome and spin-$\uparrow$ electrons can enter the dot reducing the absolute value of $S_z$.

\subsection{Heat engine for $B=0$}
\label{sec:nomag}

\begin{figure}
\includegraphics[width=\columnwidth]{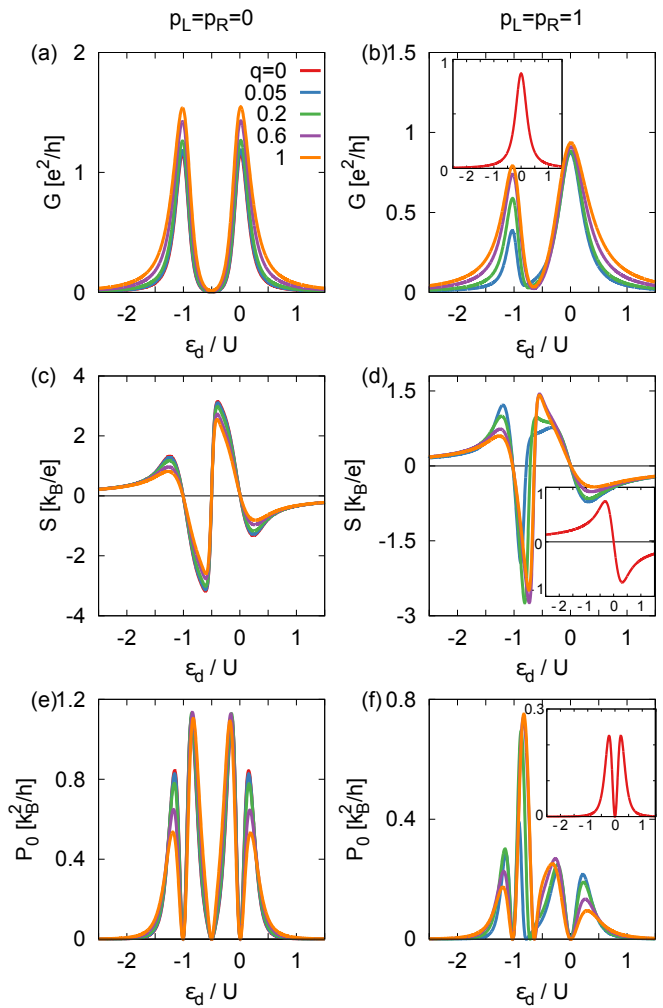}
\caption{Electrical conductance, $G$, [(a) and (b)], Seebeck coefficient, $S$, [(c) and (d)], and power factor, $P_0$, [(e) and (f)], presented as a function of the the dot's energy level, $\varepsilon_d$, and calculated in the linear response regime for indicated values of the parameter $q$ and polarization of the leads. Other parameters: $g\mu_BB=0$, $k_BT=0.5\Gamma$, and $U=10\Gamma$. Insets show the case of $q=0$ for $p_L=p_R=1$.}
\label{fig:fig3}
\end{figure}

\begin{figure}
\includegraphics[width=\columnwidth]{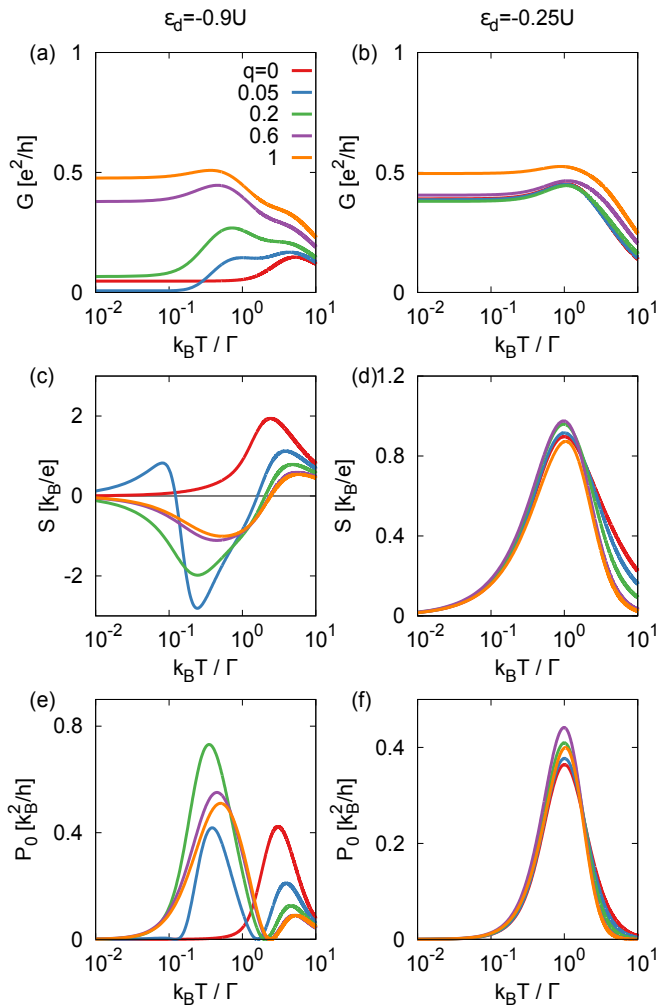}
\caption{Electrical conductance, $G$, [(a) and (b)], Seebeck coefficient, $S$, [(c) and (d)], and power factor, $P_0$, [(e) and (f)], presented as a function of temperature, $k_BT$ (in logarithmic scale), and calculated in the linear response regime for indicated values of parameter $q$ and the dot's energy level $\varepsilon_d=-0.9U$ (left column) and $\varepsilon_d=-0.25U$ (right column). Other parameters: $p_L=p_R=1$, $g\mu_BB=0$, and $U=10\Gamma$.}
\label{fig:fig4}
\end{figure}

\begin{figure*}[!ht]
\includegraphics[width=\linewidth]{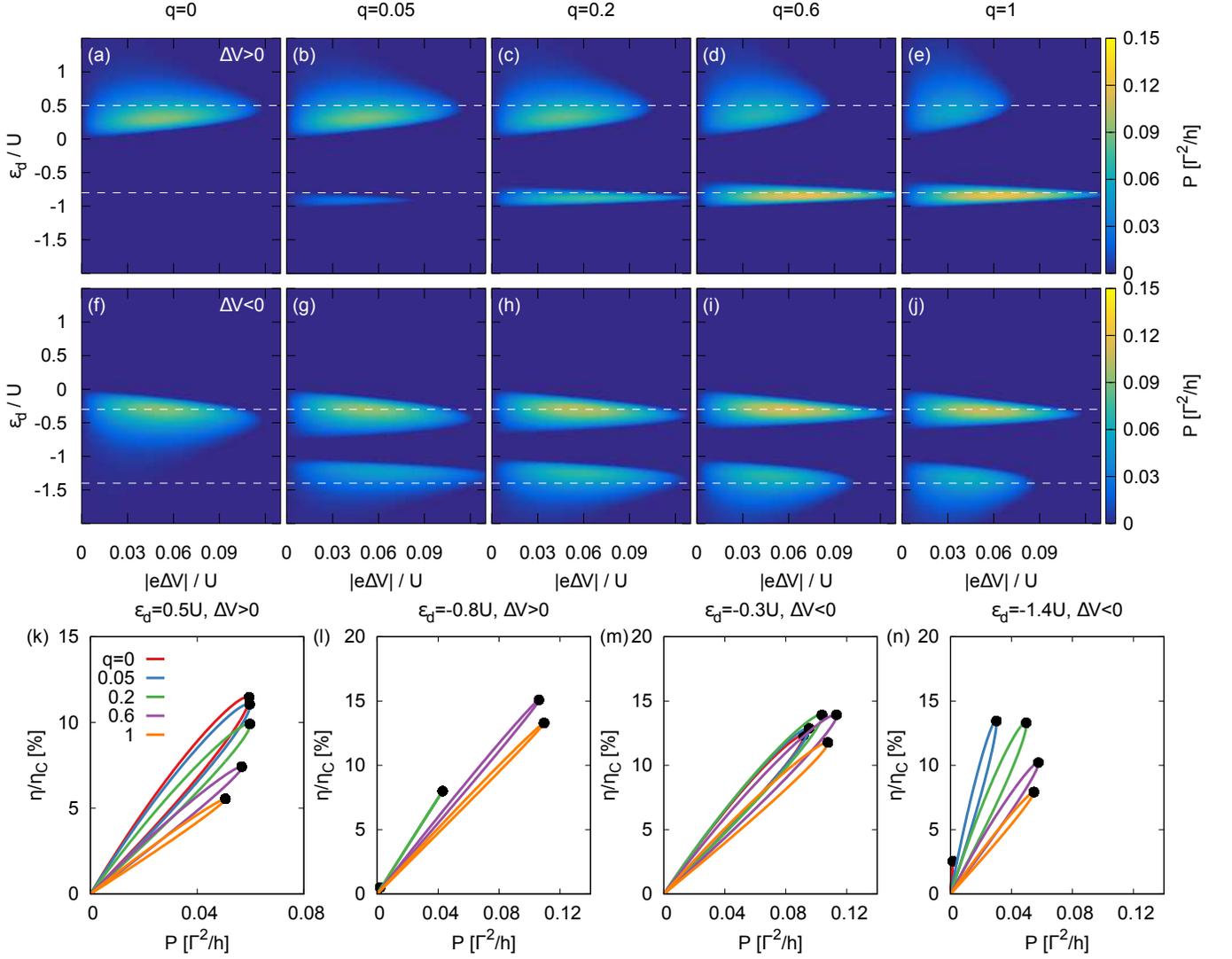}
\caption{ Power, $P$, for $\Delta V>0$ [(a)-(e)] and $\Delta V<0$ [(f)-(j)] as a function of the dot's energy level, $\varepsilon_d$, and applied bias voltage, $|e\Delta V|$, for indicated values of the parameter $q$; normalized efficiency, $\eta/\eta_C$, [(k)-(n)] as a function of the power for indicated values of the parameter $q$ and dot's energy level. Other parameters: $p_L=p_R=1$, $g\mu_BB=0$, $k_BT=0.5\Gamma$, $U=10\Gamma$, $\Delta T=2T$. White dashed lines indicate chosen energy level for the plots in [(k)-(n)]. Black dots represent the efficiency at maximal power.}
\label{fig:fig5}
\end{figure*}

In this section we show numerical results for a quantum dot-based heat engine in the absence of  external magnetic field, $B=0$. First, we analyze the electrical conductance, thermopower, and power factor in the linear response regime, and then the power and efficiency in the nonlinear regime. One should note, that thermoelectric properties of single-level quantum dots in the absence of Rashba coupling have been already investigated theoretically for different magnetic configurations of the leads, see e.g.  Ref~[\onlinecite{Swirkowicz}]. Our results on the conductance, thermopower, and power factor in a nonmagnetic case, $p_L=p_R=0$, presented in Fig.~\ref{fig:fig3}(a),~\ref{fig:fig3}(c) and~\ref{fig:fig3}(e) as a function of the dot's energy level, behave qualitatively in a similar way as the results presented in earlier works. However, they additionally show how the Rashba spin-orbit coupling modifies these parameters.

Figure~\ref{fig:fig3}(a) shows the electrical conductance as a function of the dot's energy $\varepsilon_d$ for indicated values of the Rashba spin-orbit parameter $q$.  Note that the conductance does not achieve the conductance quantum $2e^2/h$ due to a finite temperature. The two-peak structure of the conductance, i.e. the resonant peak and its Coulomb counterpart, is conserved when the Rashba coupling is nonzero.  However, both peaks become slightly broadened for $q>0$. Moreover, increase in the parameter $q$ results in a simultaneous increase of the conductance maximum due to enhanced total transmission by spin-mixing processes.

The Seebeck coefficient shown in Fig.~\ref{fig:fig3}(c) changes sign at the resonances, i.e. for $\varepsilon_d=0$ and  $\varepsilon_d=-U$, as well as in the particle-hole symmetry point, $\varepsilon_d=-U/2$. In the definition used here, the positive (negative) Seebeck coefficient corresponds to transport mediated by  holes (electrons). Thus, this figure shows that the character of transport carriers is retained for $q>0$. A weak drop in the thermopower with increasing $q$ results from increasing role of spin-mixing transmission. Due to this decrease in the  Seebeck coefficient, the power factor shown in Fig.~\ref{fig:fig3}(e) also decreases, as it is proportional to $S^2$.

In the case of half-metallic leads, $p_L=p_R=1$,  the conductance shown in Fig.~\ref{fig:fig3}(b) behaves differently. For $q=0$ there is only one peak in the conductance at $\varepsilon_d=0$, which corresponds to   tunneling of  spin-$\uparrow$ electrons through the bare dot's level. For a small nonzero  value of $q$, the peak in conductance for $\varepsilon_d=0$ is broadened, but an additional peak emerges for $\varepsilon_d=-U$. Moreover, the conductance spectrum is now asymmetric, which is typical of the Fano antiresonance. The role of resonant channel is played here by the spin-mixing channel due to Rashba spin-orbit coupling, and corresponds to spin-$\downarrow$. The background channel, in turn, corresponds to spin-$\uparrow$. The increasing rate of spin-flip processes with increasing $q$ leads to an increase in electrical conductance. It is worth noting, however, that the maximal possible value of the conductance is smaller by a factor of 2 than in the corresponding nonmagnetic case. This is because only one spin channel is available in half-metallic electrodes.

For half-metallic leads, the thermopower for $q=0$ vanishes only when $\varepsilon_d=0$, and is antisymmetric with respect to this point, as shown in Fig.~\ref{fig:fig3}(d). It acquires maximal values for $\varepsilon_d=\pm 0.2U$. In turn, the thermopower for $q>0$ is neither symmetric nor antisymmetric with respect to the particle-hole symmetry point, $\varepsilon_d=-U/2$. This results from the contribution due to spin-mixing tunneling.
Moreover, when the strength of Rashba spin-orbit coupling increases, the thermopower becomes higher. The maximal value of the thermopower is negative, which indicates particle(electron)-like character of transport. This strong dependence of the thermopower on  the type of carriers and Coulomb interaction interaction has a significant influence on the position of the dot's energy level where the Seebeck coefficient changes sign. This change occurs, as has been already discussed above, for $\varepsilon_d=0$, $U$, and $\varepsilon_\pm$, where
\begin{align}
\label{eq:eplus}
\varepsilon_{\pm}=\frac{1}{1+q}\left[1-n_\downarrow -q(1-n_\uparrow ) \right]U \hspace{2cm}\nonumber \\
 \pm\frac{1}{1+q}\sqrt{[n_\downarrow  -1  + (n_\uparrow -1) q] (n_\downarrow + n_\uparrow q) U^2}\,.
\end{align}

Since the power factor, shown in Fig.~\ref{fig:fig3}(f), reflects the structure of both thermopower and conductance, it is symmetric with respect to $\varepsilon_d=0$ for $q=0$, and  strongly asymmetric for $q>0$.  For $q>0$,  the power factor achieves the largest value for $\varepsilon_d\approx -0.75U$, where the corresponding Seebeck coefficient is maximized. It is worth noting that the  power factor for $q=0$ is significantly smaller  when compared to that in the case of $q>0$.

Temperature dependence of the linear conductance, thermopower, and power factor, shown in Fig.~\ref{fig:fig4} for two positions of the dot's energy level, $\varepsilon_d=-0.9U$ and $-0.25U$, reveals a pronounced difference between transport through the Rashba-induced and background channels. Note, $\varepsilon_d=-0.9U$ ($\varepsilon_d=-0.25U$) corresponds to the Rashba-induced (background) peak in Fig.~\ref{fig:fig3}(b).
When $q=0$, the conductance shown in Fig.~\ref{fig:fig4}(a) for $\varepsilon_d=-0.9U$ is small and reaches   a maximum at a certain value of  $k_BT>\Gamma$.
Below $k_BT =\Gamma$ it is small and constant. 
For $q>0$ the conductance displays a more complex behavior. In particular, for $q=0.05$ and low temperatures, $k_BT<0.1\Gamma$, the conductance is zero due to Fano antiresonance. When $q$ increases, position of the antiresonance with respect to $\varepsilon_d$ shifts towards lower energies, and at low temperatures the conductance  achieves a constant value.
On the other hand, for $\varepsilon_d=-0.25U$, the conductance shown in Fig.~\ref{fig:fig4}(b) varies rather weakly with increasing $q$, since the dominant contribution comes from the spin-$\uparrow$ peak which is present even for $q=0$.

Temperature dependence of the thermopower  for $\varepsilon_d=-0.9U$ and different values of the parameter $q$ is shown in Fig.~\ref{fig:fig4}(c). For $q=0$, the  thermopower is positive and small for $k_BT<\Gamma$, while for $k_BT>\Gamma$ it increases and achieves a maximal value for $k_BT\approx 2.5\Gamma$.
This peak remains for $q>0$, but its height decreases with increasing $q$ and its position shifts towards higher temperatures. Behavior of the thermopower with temperature for $q>0$ differs remarkably from that for $q=0$. This is due to the contribution from the Rashba-induced channel. Moreover, as already shown above, see Eq.~(\ref{eq:eplus}), the energy levels where the thermopower changes sign depend on the average occupation numbers which, in turn, depend on temperature. All this leads to sign reversal of the thermopower with decreasing temperature, which takes place twice for small values of $q$ [see the curve for $q=0.05$ in Fig.~\ref{fig:fig4}(c)] and once for larger values of $q$.
For $\varepsilon_d=-0.25U$ the thermopower is dominated by the background channel, and therefore it is only weakly dependent on $q$, see Fig.~\ref{fig:fig4}(d). Apart from this,
it reaches a maximal positive value at $k_BT\approx \Gamma$.

The power factor is shown in Fig.~\ref{fig:fig4}(e) for $\varepsilon_d=-0.9U$ and in Fig.~\ref{fig:fig4}(f) for $\varepsilon_d=-0.25U$. It reaches maximum values at temperatures which approximately correspond to  maximum values of the thermopower.
Interestingly, contrary to expectations, the maximum value of $P_0$ is for $q=0.2$ and not for $q=0.05$, for which the Seebeck coefficient is larger. This is due to the fact, that the conductance for $q=0.05$  in the relevant temperature range is smaller than for $q=0.2$. The maximal power factor, and thus the power generated, is larger for $\varepsilon_d=-0.9U$ than for $\varepsilon_d=-0.25U$, where the dominating contribution comes from transport through the background channel. However, lower temperatures are necessary to maximize the interference-induced $P_0$.

The key parameters that characterize a heat engine are the generated power $P$ and the efficiency $\eta$. The power generated in the system under consideration, working as a heat engine, is shown in Fig.~\ref{fig:fig5} as a function of $|e\Delta V|$ and the dot's energy level, $\varepsilon_d$, for indicated values of  $q$ and fully polarized leads, $p_L=p_R=1$. The power was calculated from Eq.~(11), in which $\Delta V$ is a voltage applied to counteract the thermally induced current $j_e$. This means that $\Delta V$ and $j_e$ have opposite signs when the system operates as a heat engine. The first row from top in Fig.~\ref{fig:fig5} corresponds to positive $\Delta V$ (negative $e\Delta V$) while the second row to negative $\Delta V$ (positive $e\Delta V$).

For $q=0$, shown in Figs.~\ref{fig:fig5}(a) and (f), the power exhibits a  single relatively broad peak, which for positive $\Delta V$ appears  for $\varepsilon_d>0$ [Fig.~\ref{fig:fig5}(a)], while for  negative $\Delta V$  appears for $\varepsilon_d<0$  [Fig.~\ref{fig:fig5}(f)].
Since the Coulomb blockade peaks are absent for $q=0$ due to full spin polarization of both leads
(double occupancy of the dot is forbidden), there is only one peak for positive and one for negative voltage.
When $\varepsilon_d>0$, then the current is dominated by electrons, while for $\varepsilon_d<0$ it is dominated by holes. Accordingly, the currents in these two regimes have opposite signs and thus the voltages against these currents also have opposite signs. Note, the corresponding Seebeck coefficients in these two regimes have opposite signs, too. The power vanishes for $\varepsilon_d= 0$,  because the corresponding Seebeck coefficient is equal to zero, so the difference in chemical potentials of the electrodes and thus also voltage disappear. The largest blocking voltage, $|eV_b|\approx 0.1U$ is obtained for $\varepsilon_d\approx 0.5U$ for positive $\Delta V$ and for $\varepsilon_d\approx -0.5U$ for negative $\Delta V$.

The suppression of double occupancy of the dot is lifted when the spin-orbit channel for transmission is open, which appears for nonzero values of $q$. The Coulomb peaks in transport characteristics  are then clearly seen, as already mentioned above. As a result, a second peak in the power appears for positive as well as for negative $\Delta V$. Indeed, for a nonzero but small $q$, an additional peak in the power  appears for $\varepsilon_d<0$ for both $\Delta V>0$ and $\Delta V<0$, as shown in Fig.~\ref{fig:fig5}(b,g) for $q=0.05$. The peak for $\Delta V>0$ and $\varepsilon_d>0$,  shown in Fig.~\ref{fig:fig5}(b), is similar to the corresponding one for $q=0$, discussed above. The corresponding maximal power, however, is lower than for $q=0$.
The additional peak appears for $-U<\varepsilon_d<\varepsilon_+$ [see Eq.~(\ref{eq:eplus})]. The corresponding maximal  power is smaller than for $\varepsilon_d>0$. In turn, for $\Delta V<0$ (and $q=0.05$), the upper peak is similar to that present for $q=0$, whereas an additional peak appears for  $\varepsilon_d$ shifted down by $U$,  as shown in Fig.~\ref{fig:fig5}(g).

Intensities of the peaks change with increasing $q$. For $\Delta V>0$, intensity of
the additional peak (absent for $q=0$) increases with increasing $q$, whereas the intensity of the peak existing at $q=0$  decreases with increasing $q$, see Figs.~\ref{fig:fig5}(a)-(e).
In turn, for $\Delta V<0$, see Figs.~\ref{fig:fig5}(f)-(j), intensity of the additional peak is rather low while the intensity of the peak existing also for $q=0$ slightly increases with increasing $q$.

It is known that for practical purposes a heat engine should work with the highest efficiency when the power is maximal. Figures~\ref{fig:fig5}(k)-(n) show the efficiency as a function of the power for indicated values of the parameter $q$ and indicated positions of the dot's energy level. These positions are indicated by white dashed lines in Figs.~\ref{fig:fig5}(a)-(j), and  correspond to the background and Rashba-induced channels. In the former case and for $\Delta V>0$, the largest efficiency is for $q=0$ and then it decreases with a further increase in $q$. In the vicinity of the interference-induced resonance, i.e. for $\varepsilon_d=-0.8U$, there is no power generated when $q=0$, and the efficiency increases non-monotonically with increasing $q$, taking the highest value for $q=0.6$. For $\Delta V<0$ and $\varepsilon_d=-0.3U$, the efficiency is roughly independent of $q$, while for $\varepsilon_d\approx -1.3U$ it decreases monotonically with increasing $q$. The maximal efficiency for both voltage polarities, however, is lower than the appropriate Carnot efficiency and lower than the Curzon-Ahlborn efficiency, $\eta_{\text{CA}}/\eta_{\text{C}}\approx 0.64$.

\subsection{Heat engine for $B\neq 0$}
\label{sec:mag}

\begin{figure}
\includegraphics[width=\columnwidth]{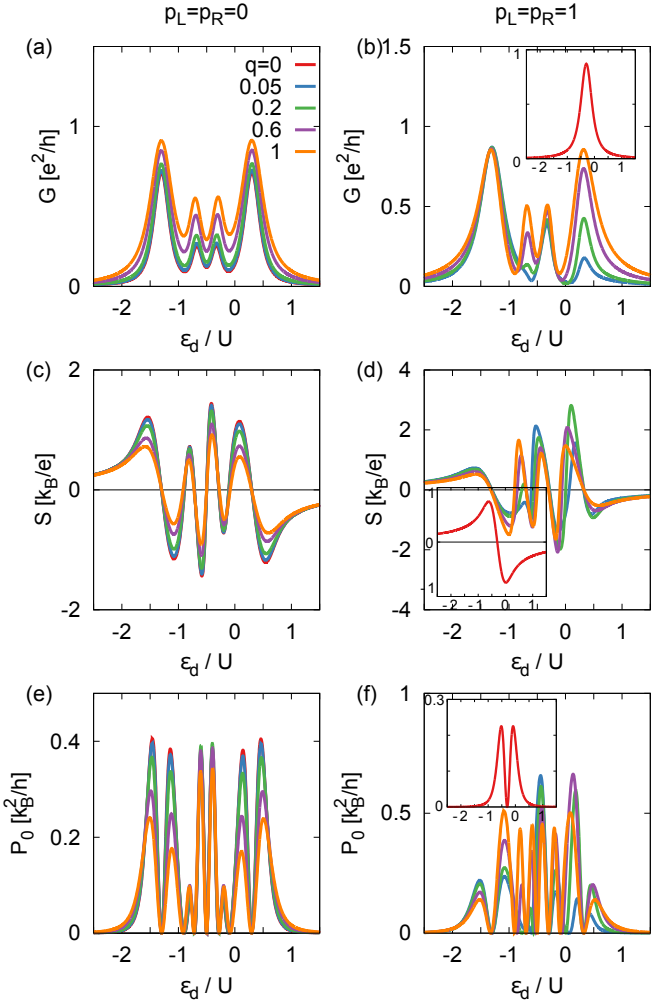}
\caption{Electrical conductance, $G$, [(a) and (b)], Seebeck coefficient, $S$, [(c) and (d)], and power factor, $P_0$, [(e) and (f)], presented as a function of the the dot's bare energy level, $\varepsilon_d$, and calculated in the linear response regime for indicated values of the parameter $q$ and polarization of the leads. Other parameters: $g\mu_BB=0.6U$, $k_BT=0.5\Gamma$, and $U=10\Gamma$. Insets show the case of $q=0$ for $p_L=p_R=1$.}
\label{fig:fig6}
\end{figure}

\begin{figure}
\includegraphics[width=\columnwidth]{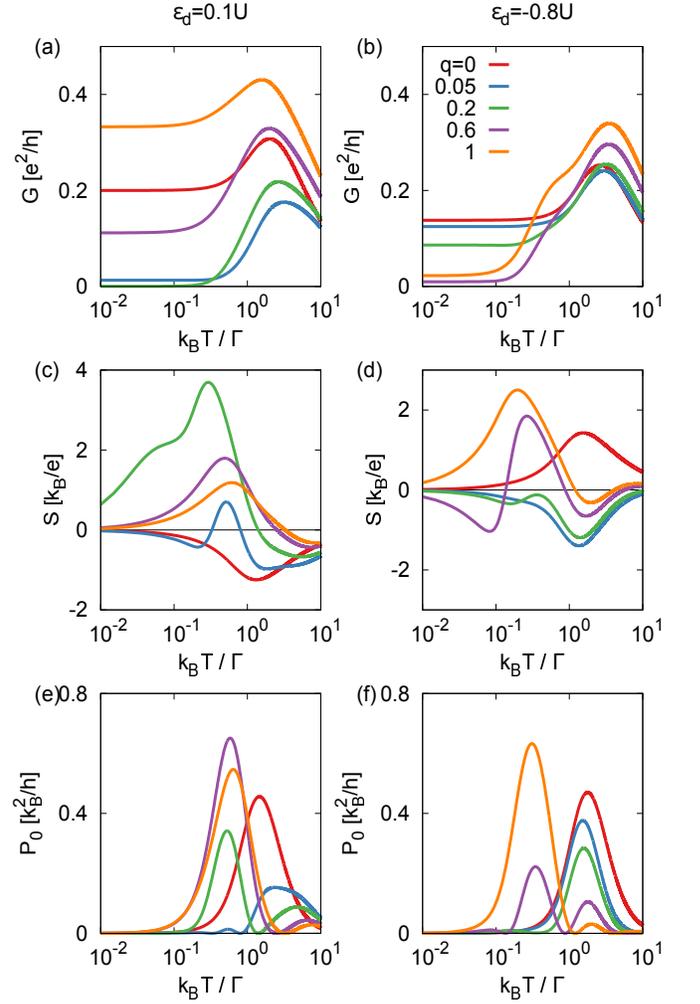}
\caption{Electrical conductance, $G$, [(a) and (b)], Seebeck coefficient, $S$, [(c) and (d)], and power factor, $P_0$, [(e) and (f)], presented as a function of temperature, $k_BT$ (in logarithmic scale), and calculated in the linear response regime for indicated values of the parameter $q$, and for $\varepsilon_d=0.1U$ (left column) and $\varepsilon_d=-0.8U$ (right column). Other parameters: $p_L=p_R=1$, $g\mu_BB=0.6U$, $k_BT=0.5\Gamma$, and $U=10\Gamma$.}
\label{fig:fig7}
\end{figure}
\begin{figure*}
\includegraphics[width=\linewidth]{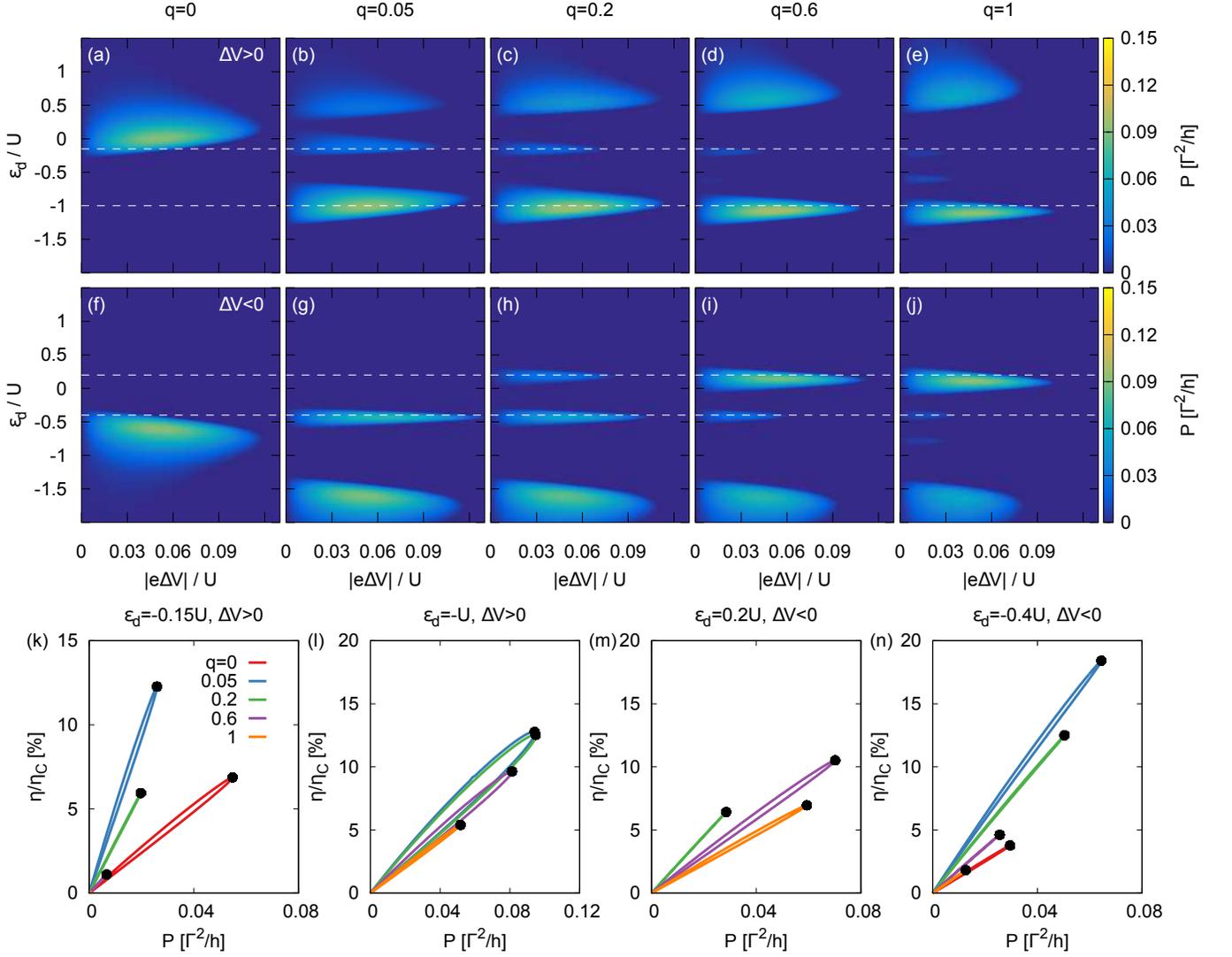}
\caption{ Power, $P$, [(a)-(j)] as functions of the dot's energy level, $\varepsilon_d$, and applied bias voltage, $|e|\Delta V$, calculated for indicated values of the parameter $q$ and $\Delta V >0$ [(a)-(e)] and $\Delta V <0$ [(f)-(j)]. Normalized efficiency [(k)-(n)] as a function of the power for indicated values of the parameter $q$ and dot's energy level. Other parameters: $p_L=p_R=1$, $g\mu_BB=0.6U$, $k_BT=0.5\Gamma$, $U=10\Gamma$, and $\Delta T=2T$. White dashed lines indicate chosen energy levels for the plots in [(k)-(n)]. Black dots in [(k)-(n)] represent the efficiency at maximal power.}
\label{fig:fig8}
\end{figure*}

For  $B\neq 0$ and non-magnetic leads, $p_L=p_R=0$, the electrical conductance shown in Fig.~\ref{fig:fig6}(a) as a function of the dot's energy level $\varepsilon_d$ displays two additional peaks due to a relatively large Zeeman splitting, $g\mu_BB=0.6U$. The conductance is symmetric with respect to the particle-hole symmetry point, $\varepsilon_d=-U/2$.  The increase in $q$ leads to an increase in conductance due to the spin-mixing processes already discussed in the previous section.
In turn, the  thermopower shown in Fig.~\ref{fig:fig6}(c) changes sign two times more than in the case of $B=0$. Such a behavior is typical for multilevel system, so we will not discuss it in more detail.
With increasing $q$, the points where the Seebeck coefficient vanishes are slightly shifted away from the points  where $S$ vanishes for $q=0$, except  the particle-hole symmetry point which is preserved.
The corresponding power factor is shown in Fig.~\ref{fig:fig6}(e). Since the number of the points where the thermopower vanishes is now larger, the dependence of the power factor on the dot's energy level is more complex, i.e. the number of peaks is larger.  Heights of these peaks, however, decrease with increasing parameter $q$.

The conductance, thermopower, and power factor in the case of half-metallic leads are shown in Figs.~\ref{fig:fig6}(b),~\ref{fig:fig6}(d), and~\ref{fig:fig6}(f).
Since there is only one spin channel in the leads, only one component of the Zeeman-split level is active in transport for $q>0$, and therefore only one peak appears in the conductance when Rashba coupling vanishes.
Because energy of the spin-$\uparrow$ level is shifted up by $g\mu_BB/2$ due to Zeeman energy, the corresponding peak appears for $\varepsilon_{d}=-g\mu_BB/2$.  However,  a more complex conductance spectrum emerges when $q>0$. First, both components of the Zeeman-split level contribute to transport. Second, the Coulomb counterparts also appear as now two electrons of opposite spins can reside in the dot. Due to the Zeeman splitting, the background channels appear at $\varepsilon_{d}=-g\mu_BB/2$ and $\varepsilon_{d}=-U-g\mu_BB/2$, where changes in $q$ do not lead to significant qualitative and quantitative modifications of the conductance. However, for $\varepsilon_d=g\mu_BB/2$ and $-U+g\mu_BB/2$ the conductance strongly depends on $q$.
For $q=0.05$ a sharp antiresonance develops in the conductance due to destructive interference.
In turn, the corresponding thermopower for $q=0$ changes sign only for $\varepsilon_d=\varepsilon_\uparrow$, as follows from the inset in Fig.~\ref{fig:fig6}(d). For $q>0$, behavior of the thermopower with the dot's energy level is more complex and is correlated with the corresponding conductance spectra, as already discussed before.
The corresponding
power factor, shown in Fig.~\ref{fig:fig6}(f) is relatively low and for $q=0$ is symmetric with respect to $\varepsilon_d -g\mu_BB/2=\varepsilon_\uparrow$. The power factor becomes highly asymmetric for $0<q<1$.
The highest values obtained strongly depend on $q$  and are smaller by a factor of about 2 from  the corresponding maximal power factor obtained in the case of nonmagnetic leads.

\begin{figure*}
\includegraphics[width=\linewidth]{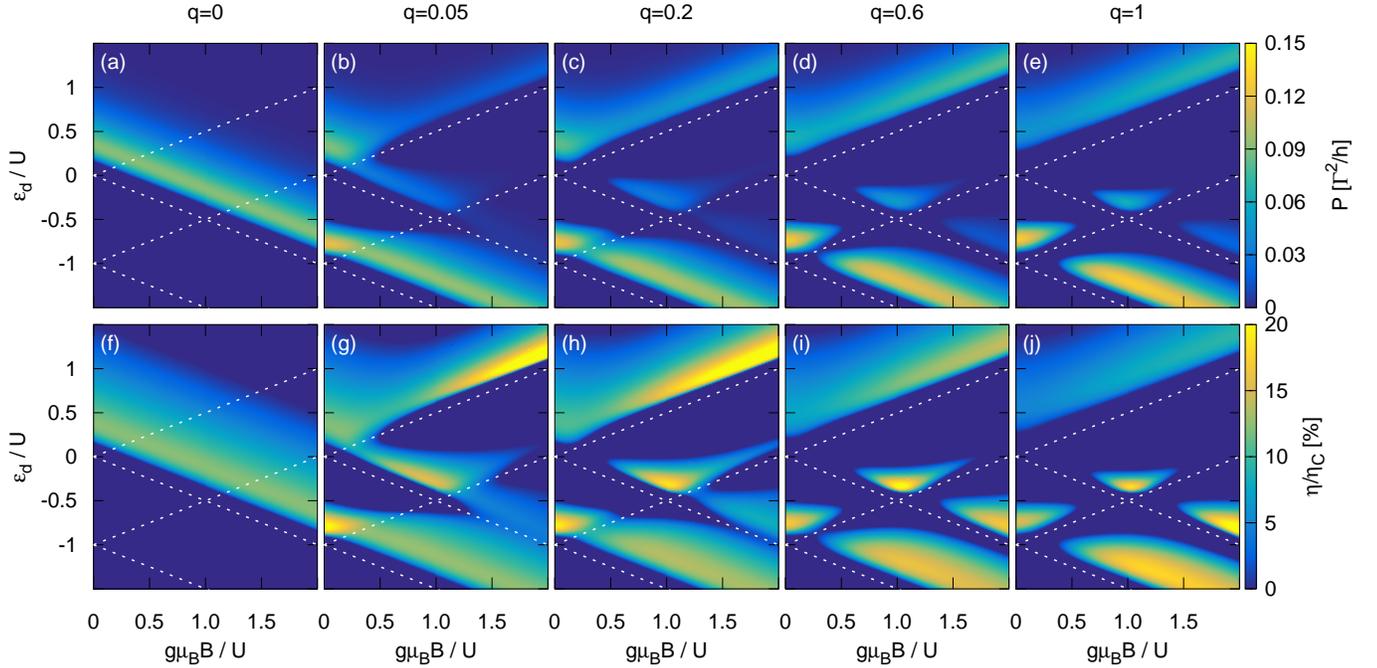}
\caption{ Power, $P$, [(a)-(d)], and normalized efficiency, $\eta/\eta_C$, [(e)-(h)], as functions of the dot's energy level, $\varepsilon_d$, and magnetic field, $g\mu_BB$, for indicated values of parameter $q$. Other parameters: $p_L=p_R=1$, $k_BT=0.5\Gamma$, $U=10\Gamma$, $e\Delta V=-0.05U$, and $\Delta T=2T$. White dotted lines represent spin-dependent energy levels, $\varepsilon_\uparrow$ and $\varepsilon_\downarrow$.}
\label{fig:fig9}
\end{figure*}

Due to the Zeeman splitting of the dot's level,
the temperature dependence of electric and thermoelectric coefficients is more complex than in the absence of magnetic field.
This temperature dependence of electric conductance, thermopower, and power factor for $p_L=p_R=1$ is shown in Fig.~\ref{fig:fig7} for indicated values of $q$ and two values of the dot's energy level, $\varepsilon_d=0.1U$ and $\varepsilon_d=-0.8U$.
From Fig.~\ref{fig:fig6}(b) follows that both these energies are in the vicinity of the peaks associated with Rashba-induced channels, where the dependence on $q$ is quite significant. This leads to a nontrivial dependence of the electric conductance, thermopower, and power factor on temperature. Physical origin of this behavior is similar to that presented already in the case of zero magnetic field, so we will not describe it in more details. The only difference follows from the Zeeman splitting due to external magnetic field.

Finite Zeeman splitting of the dot's energy level leads to additional peaks in power -- similarly as it leads to  the additional peaks in other transport/thermoelectric quantities discussed above. The power generated in the system under consideration in the presence of  external magnetic field is  shown in Fig.~\ref{fig:fig8}(a)-(j). The upper row in this figure [(a)-(e)] corresponds to $\Delta V >0$, while the second row [(f)-(j)] corresponds to $\Delta V<0$.

For $q=0$, the power spectrum shown in Fig.~\ref{fig:fig8}(a) and Fig.~\ref{fig:fig8}(f) for $\Delta V>0$ and $\Delta V<0$, respectively, is similar to the corresponding one in the absence of external magnetic field, i.e. it exhibits only a single peak due to a single active spin channel only, which, however, is shifted towards lower bare dot's energy level. This is because transport goes  through the $\varepsilon_\uparrow$ component of the Zeeman spin-split dot's level. The blocking voltage, $|eV_b|\approx 0.1U$, is similar to that for zero magnetic field and is generally greater than that for $q>0$.

As in the case of zero magnetic field,  the Rashba spin-flip tunneling for $q>0$ opens the second spin channel for electronic transport through the quantum dot. Moreover, it also leads to interference effects (Fano antiresonance). The power generated for $q=0.05$ is shown in Figs. ~\ref{fig:fig8}(b) for $\Delta V>0$ and \ref{fig:fig8}(g) for $\Delta V<0$. From these figures follows that the number of peaks in the power as a function of the bare dot's level energy $\varepsilon_d$ is larger. Generally, one or two additional peaks are well resolved for both positive and negative $\Delta V$ when $q>0$. Their position  is well correlated with maxima of the thermopower shown in Fig.~\ref{fig:fig6}. The increased number of peaks follows from opening of the double occupancy of the dot by spin-flip tunneling, and from Zeeman splitting of the dot's level. For $\Delta V>0$,  the  maximal power is associated with the lowest peak at $\varepsilon_d/U\approx 1$. The peaks corresponding to the dot's energy above the Fermi level, $\varepsilon_d>0$, are less pronounced. In particular, the power vanishes for $\varepsilon_d\approx g\mu_BB/2$. The blocking voltages for $\varepsilon_d\approx 0$ and  $\varepsilon_d\approx 0.5U$ are lower than the corresponding values for $q=0$.
For $q>0.05$, the maximal power and blocking voltage for the background channel, i.e. for $\varepsilon_d>0.5U$ and for $\varepsilon_d<-1.5U$, decrease with increasing $q$ for both $\Delta V>0$ and $\Delta V<0$, while the maximal power of the spin-flip-induced channels increases with $q$.

Efficiency as a function of the power is shown in Figs.~\ref{fig:fig8}(k)-(m) for selected positions of the dot's energy level. For $\varepsilon_d=-0.15U$ and $\Delta V>0$, there is no power generated for $q=0$, as also follows from  Fig.~\ref{fig:fig8}(f). The highest efficiency at maximal power is achieved for $q=0.05$, which results from the Fano antiresonance, as already described above. Similar behavior is observed for $\varepsilon_d=-0.4U$ and $\Delta V<0$. For higher values of $q$, the efficiency and efficiency at maximal power quickly decrease. For $\Delta V>0$ and $\varepsilon_d=-U$, the  efficiency is roughly constant for small $q$ and decreases with increasing $q$. For $\Delta V<0$ and $\varepsilon_d=0.2U$, it is maximal for $q=0.6$.

The power generated in the system as well as the corresponding efficiency remarkably depend on the strength of  magnetic field, as shown in Fig.~\ref{fig:fig9} for the bias voltage roughly corresponding to maximal values of the power and efficiency shown in Fig.~\ref{fig:fig8}, i.e. $|e|\Delta V=0.05U$. For $q=0$, the power and efficiency shown in Figs.~\ref{fig:fig9}(a) and (f), are roughly constant with respect to magnetic field. However, the dot's energy level, for which the heat-to-work conversion occurs, depends linearly on magnetic field, as the transport occurs through the single spin-$\uparrow$ level.

For $q>0$, the spin-$\downarrow$ channel is activated due to finite Rashba spin-orbit coupling. This activation leads to complex interplay between the effects due to external magnetic field, Rashba spin-orbit coupling, and Coulomb interaction. This, in turn, results in additional peaks in both power shown in Fig.~\ref{fig:fig9}(b) for $q=0.05$ and efficiency shown in Fig.~\ref{fig:fig9}(g) also for $q=0.05$. The highest power output is generated for $\varepsilon_d\approx -0.75U$ and relatively small magnetic fields,  $g\mu_BB<0.25U$.
For  $g\mu_BB=U$, energy of spin-$\uparrow$ and spin-$\downarrow$ levels corresponding to the $\varepsilon_d=0$ and $\varepsilon_d=-U$ resonances are equal, so the spin-$\uparrow$ in the dot is flipped to spin-$\downarrow$ for $0>\varepsilon_d>-U/2$, i.e. when a second electron overcomes Coulomb blockade and enters the dot. Moreover the maximal efficiency increases for $\varepsilon_d>0.5U$ and $g\mu_BB>U$.

For $q=0.2$, the power shown in Figs.~\ref{fig:fig9}(c) is not generated for $\varepsilon_d=g\mu_BB/2$ and $\varepsilon_d=-g\mu_BB/2$, and the efficiency corresponding to these energy levels, shown in Fig.~\ref{fig:fig9}(h), vanishes. However, due to the Fano antiresonance, there is a finite power and efficiency for resonant energy of spin-$\uparrow$ level, i.e. for $\varepsilon_d=-U+g\mu_B/2$ and magnetic field $U/4<g\mu_BB<U/2$ or  $1.25U<g\mu_BB<1.75U$. The maximal efficiency is obtained for $g\mu_BB>U/2$ and $\varepsilon_d>g\mu_BB/2$.

When $q>0.2$, the power and efficiency shown in Figs.~\ref{fig:fig9}[(d),(e)] and Figs.~\ref{fig:fig9}[(i),(j)], respectively, vanish for $\varepsilon_d=\pm g\mu_BB/2$ and for $\varepsilon_d=-U\pm g\mu_BB/2$, since these energies are well separated and the Fano-like antiresonance, present for lower $q$, does not contribute to transport. Maximal efficiency can be obtained for large magnetic fields, i.e. for $g\mu_BB>1.5U$ and $\varepsilon_d<-U/2$, i.e. when
transport occurs mainly through the $\varepsilon_\downarrow$ level.

\section{Summary}
\label{sec:concl}

In conclusion, we have analyzed a heat engine based on a quantum dot connected to half-metallic ferromagnetic leads.  Both, spin-conserving and spin-flip tunneling processes between the dot and electrodes were taken into account. The latter occur
due to Rashba spin-orbit interaction.
Basic parameters of the engine, such as power and efficiency, can be then modulated not only by properties of the dot itself, such as position of the energy level or Coulomb correlation parameter, but also by Rashba spin-orbit coupling. The  latter, in principle, can be controlled by external electric field.

A particularly interesting case occurs when the ferromagnetic leads are fully spin-polarized and have aligned magnetic moments (parallel configuration).
Even though only electrons of one spin orientation are then present in the electrodes, the Rashba spin-orbit coupling activates the dot's level of the opposite spin, so the effects due to double occupancy (Coulomb blockade) play an important role. Moreover, this also leads to resonant effects, especially to the Fano-like interference, where the spin-$\downarrow$ channel takes the role of a resonant channel, while the spin-$\uparrow$ channel assumes the role of background channel. This, in turn, leads to an enhanced thermoelectric response of the system. Moreover, when an external magnetic field is applied to the system, the complex interplay between the effects due to spin-orbit coupling, magnetic field, and Coulomb interaction leads to
spin-selective power generation.

\end{document}